\documentstyle[12pt,epsf]{article}

\renewcommand{\bar}[1]{\overline{#1}}
\newcommand{\M}{{\cal M}}
\newcommand{\VEV}[1]{\left\langle{#1}\right\rangle}
\newcommand{\etal}{{\em et al.}}
\newcommand{\ie}{{\it i.e.}}
\newcommand{\eg}{{\it e.g.}}

\newcommand{\ket}[1]{\vert\,{#1}\rangle}

\begin{document}
\begin{flushright}
SLAC--PUB--8295 \\
November 1999
\end{flushright}
\bigskip\bigskip

\centerline{{QCD PROCESSES AT THE AMPLITUDE LEVEL}
    \footnote{\baselineskip=14pt
     Work supported by the Department of Energy, contract 
     DE--AC03--76SF00515.}}
\vspace{22pt}
  \centerline{\bf Stanley J. Brodsky}
\vspace{8pt}
  \centerline{\it Stanford Linear Accelerator Center}
  \centerline{\it Stanford University, Stanford, California 94309}
  \centerline{e-mail: sjbth@slac.stanford.edu}
\vspace*{0.9cm}
\begin{center}
Abstract
\end{center}
Some of the most difficult theoretical challenges of
QCD occur  at the interface of
the perturbative and nonperturbative regimes.  Exclusive and semi-exclusive
processes,  the diffractive dissociation of hadrons into jets, and hard
diffractive processes such as vector meson leptoproduction provide new testing
grounds for QCD and essential information on the structure of light-cone
wavefunctions of hadrons, particularly the pion distribution amplitude.  I
review
the basic features of the leading-twist QCD predictions and the problems and
challenges of studying QCD at the amplitude level.

\vfill
\begin{center}
Presented at the TJNAF Workshop \\
``The Transition from Low to High Momentum Transfer Form Factors" \\
 September 17,  1999 \\
 The University of Georgia, Athens, Georgia
\end{center}
\vfill
\newpage

\section{Introduction}
An audacious claim of the Standard Model is the assertion that
all of the properties of hadrons, all of their strong interactions, and even
nuclear physics can be derived from just one line, the Lagrangian density of
quantum chromodynamics. This assertion is even more remarkable considering that
the fundamental quanta and the color charges of QCD cannot
be directly observed.  In addition to the set of quark masses,
$\{m_f\}$ the only dynamical mass scale of QCD is $\Lambda_{QCD} \sim 200$ GeV,
the momentum scale where the $QCD$ coupling becomes large.

The traditional focus of theoretical work in QCD has been on hard
inclusive processes and jet physics where perturbative methods and
leading-twist factorization provide predictions up to next-to-next-to
leading order. Most of these predictions appear to be validated by
experiment with good precision. More recently, the
domain of reliable perturbative QCD predictions has been extended to
much more complex phenomena, such as the BFKL approach to the hard QCD pomeron
in deep inelastic scattering at small $x_{bj}$,
\cite{Balitsky:1978ic} virtual photon scattering,\cite{Brodsky:1997sd}
and the energy dependence of hard virtual photon diffractive
processes, such as $\gamma^* p \to
\rho^0 p$.\cite{BGMFS}

Now a primary  goal of both high
energy and nuclear physics is to unravel the nonperturbative structure and
dynamics of nucleons and nuclei in terms of their fundamental quark and gluon
degrees of freedom. There are many applications of QCD where the
non-perturbative composition of hadrons in terms of their quark and
gluon degrees of freedom play a crucial role, for example the
$x_{bj}$-dependence of structure functions measured in deep inelastic
scattering, exclusive and semi-exclusive processes such as form factors,
two-photon processes, elastic scattering at fixed
$\theta_{cm}$, as well as the semi-leptonic decays of heavy hadrons.  The
analysis of QCD processes at the amplitude level is a challenging
problem, mixing issues involving non-perturbative and perturbative
dynamics.

Deep inelastic lepton-proton scattering has provided the traditional guide to
the proton's structure. When the photon virtuality is of order of the quark
intrinsic transverse momentum, evolution from QCD radiative processes becomes
quenched, and the structure functions reveal fundamental features of the
proton's
composition. The data in fact indicate a nonperturbative structure of
nucleons more complex than a simple three quark bound state.  For example,
if the
sea quarks were generated solely by perturbative QCD evolution
via gluon splitting, the anti-quark distributions would be approximately isospin
symmetric. However, the $\bar u(x)$ and
$\bar d(x)$ antiquark distributions of the proton at $Q^2
\sim 10$ GeV$^2$ are found to be quite different in
shape \cite{{Nasalski:1994bh}}
and thus must reflect dynamics intrinsic to the proton's structure.
Evidence for a difference between the $\bar s(x)$ and $s(x)$ distributions
has also been claimed. \cite{Barone:1999yv}

It is helpful to categorize the parton distributions as ``intrinsic"---pertaining to
the long-time scale composition of the target hadron, and ``extrinsic",---reflecting 
the short-time substructure of the individual quarks and gluons themselves.
Gluons carry a
significant fraction of the proton's spin as well as its momentum.  Since
gluon exchange between valence quarks contributes to the
$p-\Delta$ mass splitting, it follows that the gluon distributions 
cannot be solely accounted for by gluon bremsstrahlung from
individual quarks, the
process responsible for DGLAP evolutions of the structure functions.
Similarly,  in the case of heavy quarks, $s\bar s$,
$c \bar c$, $b \bar b$, the diagrams in which the sea quarks are multiply
connected to
the valence quarks are intrinsic to the proton structure itself. \cite{IC}
The $x$ distribution of intrinsic heavy quarks is peaked at large $x$
reflecting the fact that higher Fock state wavefunctions
containing heavy quarks are maximal when the off-shellness of
the fluctuation is minimized.  The evidence for intrinsic charm at large $x$
in deep inelastic scattering is discussed by Harris \etal  \cite{Harris:1996jx}
Thus neither gluons nor sea
quarks are solely generated by DGLAP evolution, and one cannot define a
resolution scale $Q_0$ where the sea or
gluon degrees of freedom can be neglected.

There have also been surprises associated with the chirality distributions
$\Delta q = q_{\uparrow/\uparrow} - q_{\downarrow/\uparrow}$ of the valence
quarks which show that a simple valence quark
approximation to nucleon spin structure functions is far from the actual
dynamical situation.\cite{Karliner:1999fn}

Part of the complexity of hadronic physics is related to the fact that a
relativistic bound state of a quantum field theory fluctuates not only in
momentum space and helicity, but also in particle number.  For example, the
heavy quark sea is associated with higher particle number states.  Fortunately
we can use the light-cone Fock expansion to provide a frame-independent
representation of a hadron in terms of a set of wavefunctions
$\{\psi_{n/H}(x_i,\vec k_{\perp i},\lambda_i)\}$ describing its
composition into relativistic quark and gluon constituents.
The light-cone Fock
representation of QCD obtained by quantizing the theory at fixed ``light-cone"
time
$\tau = t+z/c$.\cite{PinskyPauli} This representation is the
extension of Schr\"odinger many-body theory to the relativistic domain.
For example, the proton state has the Fock expansion
\begin{eqnarray}
\ket p &=& \sum_n \VEV{n\,|\,p}\, \ket n \nonumber \\
&=& \psi^{(\Lambda)}_{3q/p} (x_i,\vec k_{\perp i},\lambda_i)\,
\ket{uud} \\[1ex]
&&+ \psi^{(\Lambda)}_{3qg/p}(x_i,\vec k_{\perp i},\lambda_i)\,
\ket{uudg} + \cdots \nonumber
\label{eq:b}
\end{eqnarray}
representing the expansion of the exact QCD eigenstate on a non-interacting
quark and gluon basis.  The probability amplitude
for each such
$n$-particle state of on-mass shell quarks and gluons in a hadron is given by a
light-cone Fock state wavefunction
$\psi_{n/H}(x_i,\vec k_{\perp i},\lambda_i)$, where the constituents have
longitudinal light-cone momentum fractions
$
x_i ={k^+_i}/{p^+} = (k^0_i+k^z_i)/(p^0+p^z)\ , \sum^n_{i=1} x_i= 1
$,
relative transverse momentum
$\vec k_{\perp i} \ , \sum^n_{i=1}\vec k_{\perp i} = \vec 0_\perp$,
and helicities $\lambda_i.$ The effective lifetime of each configuration
in the laboratory frame is ${2 P_{lab}/({\M}_n^2- M_p^2}) $ where
$
\M^2_n = \sum^n_{i=1}(k^2_{\perp i} + m^2_i)/x_i < \Lambda^2 $
is the off-shell invariant mass and $\Lambda$ is a global
ultraviolet regulator.

A crucial feature of the light-cone formalism is
the fact that the form of the
$\psi^{(\Lambda)}_{n/H}(x_i,
\vec k_{\perp i},\lambda_i)$ is invariant under longitudinal boosts; \ie,\ the
light-cone wavefunctions expressed in the relative coordinates $x_i$ and
$k_{\perp i}$ are independent of the total momentum
$P^+$,
$\vec P_\perp$ of the hadron.
The ensemble
\{$\psi_{n/H}$\} of such light-cone Fock
wavefunctions is a key concept for hadronic physics, providing a conceptual
basis for representing physical hadrons (and also nuclei) in terms of their
fundamental quark and gluon degrees of freedom.  Each Fock state interacts
distinctly; \eg, Fock states with small particle number and small impact
separation have small color dipole moments and can traverse a nucleus with
minimal interactions.  This is the basis for the predictions for ``color
transparency". \cite{BM}

Given the
$\psi^{(\Lambda)}_{n/H},$ we can construct any spacelike electromagnetic or
electroweak form factor from the diagonal overlap of the LC
wavefunctions.\cite{BD} The natural formalism for
describing the hadronic wavefunctions which enter exclusive and
diffractive amplitudes is the light-cone expansion.
Similarly, the matrix elements of the currents that define quark and gluon
structure functions can be computed from the integrated squares of the LC
wavefunctions.\cite{LB}

Can we ever hope to predict the light-cone wavefunctions from first principles
in QCD?  In the Discretized Light-Cone Quantization (DLCQ) method,\cite{DLCQ}
periodic boundary conditions are introduced in order to render the set of
light-cone momenta
$k^+_i, k_{\perp i}$ discrete.  Solving QCD then becomes reduced to
diagonalizing the mass operator of the theory.  Virtually any
$1+1$ quantum field theory, including ``reduced QCD" (which has both quark and
gluonic degrees of freedom) can be completely solved using
DLCQ.\cite{Kleb,AD}  The method yields not only the bound-state and continuum
spectrum, but also the light-cone wavefunction for each eigensolution.  The
method is particularly elegant in the case of supersymmetric theories. \cite{Antonuccio:1999ia}
 The
solutions for the model 1+1 theories can provide an important theoretical
laboratory for testing approximations and QCD-based models.  Recent progress in
DLCQ has been obtained for
$3+1$ theories utilizing Pauli-Villars ghost fields to provide a covariant
regularization.  Broken supersymmetry may be the key method for regulating
non-Abelian theories.  Light-cone gauge allows one to utilize only the
physical degrees of freedom of the gluon field.  However,  light-cone
quantization in Feynman gauge has a number of attractive features, including
manifest covariance and a straightforward passage to the Coulomb limit in the
case of static quarks.\cite{Srivastava:1999gi}

Exclusive hard-scattering reactions and hard diffractive reactions are now
providing
an invaluable window into the structure and
dynamics of hadronic amplitudes.  Recent measurements of the
photon-to-pion transition form factor at CLEO,\cite{Gronberg:1998fj} the
diffractive dissociation of pions into jets at Fermilab,\cite{E791}
diffractive vector meson leptoproduction at Fermilab and HERA, and the new
program
of experiments on exclusive proton and deuteron processes at Jefferson
Laboratory
are now yielding fundamental information on hadronic wavefunctions,
particularly the
distribution amplitude of mesons.  There is now strong evidence for color
transparency from such processes.  Such information is also critical for
interpreting exclusive heavy hadron decays and the matrix elements and
amplitudes
entering $CP$-violating processes at the $B$ factories.

In addition to the light-cone expansion, a number of theoretical tools are
available:

1.  Factorization theorems for hard exclusive, semi-exclusive, and
diffractive processes allow a rigorous separation of soft
non-perturbative dynamics of the bound state hadrons from the hard
dynamics of a perturbatively-calculable quark-gluon scattering
amplitude.  The key non-perturbative input is the
gauge and frame independent hadron distribution amplitude \cite{LB}
defined as the integral over transverse momenta of the valence (lowest
particle number) Fock wavefunction; \eg\ for the pion
\begin{equation}
\phi_\pi (x_i,Q) \equiv \int d^2k_\perp\, \psi^{(Q)}_{q\bar q/\pi}
(x_i, \vec k_{\perp i},\lambda)
\label{eq:f1}
\end{equation}
where the global cutoff $\Lambda$ is identified with the resolution $Q$.
The distribution amplitude controls leading-twist exclusive amplitudes
at high momentum transfer, and it can be related to the gauge-invariant
Bethe-Salpeter wavefunction at equal light-cone time $\tau = x^+$.

2.  The logarithmic evolution of hadron distribution amplitudes
$\phi_H (x_i,Q)$ can be derived from the perturbatively-computable tail
of the valence light-cone wavefunction in the high transverse momentum
regime.\cite{LB}

3.  Conformal symmetry provides a template for
QCD predictions, leading to relations between observables which are
present even in a theory which is not scale invariant.  For example, the natural
representation of distribution amplitudes is in terms of an expansion
of orthonormal conformal functions multiplied by anomalous dimensions
determined by
QCD evolution equations.\cite{Brodsky:1980ny,Muller:1994hg}
Thus an important guide in QCD
analyses is to identify the underlying conformal relations of QCD which are
manifest if we drop quark masses and effects due to the running of the QCD
couplings.  In fact, if QCD has an infrared fixed point (vanishing of the Gell
Mann-Low function at low momenta), the theory will closely resemble a scale-free
conformally symmetric theory in many applications.

4.  Commensurate scale relations\cite{Brodsky:1995eh} are perturbative QCD
predictions which relate observable to observable at fixed relative
scale, such as the ``generalized Crewther relation",\cite{Brodsky:1996tb} which
connects the Bjorken and Gross-Llewellyn Smith deep inelastic scattering sum
rules to measurements of the $e^+ e^-$ annihilation cross section.  The
relations
have no renormalization scale or scheme ambiguity.  The coefficients in
the perturbative series for commensurate scale relations are identical to
those of conformal QCD; thus no infrared renormalons are
present.\cite{Brodsky:1999gm} One can identify
the required conformal coefficients at any finite order by
expanding the coefficients of the usual PQCD expansion around a formal
infrared fixed point, as in the Banks-Zak method.\cite{BGGR}  All non-conformal
effects are absorbed by fixing the ratio of the respective momentum transfer and
energy scales.  In the case of fixed-point theories, commensurate scale
relations
relate
both the ratio of couplings and the ratio of scales as the fixed point is
approached.\cite{Brodsky:1999gm}

5.  $\alpha_V$ Scheme.  A natural scheme for defining the QCD
coupling in exclusive and other processes is the $\alpha_V(Q^2)$ scheme defined
from the potential of static heavy quarks.  Heavy-quark lattice gauge theory
can provide highly precise values for the coupling.  All vacuum polarization
corrections due to fermion pairs are then automatically and analytically
incorporated into the Gell Mann-Low function, thus avoiding the problem of
explicitly computing and resumming quark mass corrections related to the running
of the coupling.  The use of a finite effective charge such as
$\alpha_V$ as the expansion parameter also provides a basis for regulating the
infrared nonperturbative domain of the QCD coupling.

6.  The Abelian Correspondence Principle.  One can consider QCD
predictions as analytic functions of the number of colors $N_C$ and
flavors $N_F$.  In particular, one can show at all orders of
perturbation theory that PQCD predictions reduce to those of an Abelian
theory at $N_C \to 0$ with ${\widehat \alpha} = C_F \alpha_s$ and
${\widehat N_F} = N_F/T C_F$ held fixed.\cite{Brodsky:1997jk} There is
thus a deep connection between QCD processes and their corresponding QED
analogs.

\section{Electoweak Decays and the Light-Cone Fock Expansion}

\begin{figure}[htb]
\begin{center}
\leavevmode
\epsfbox{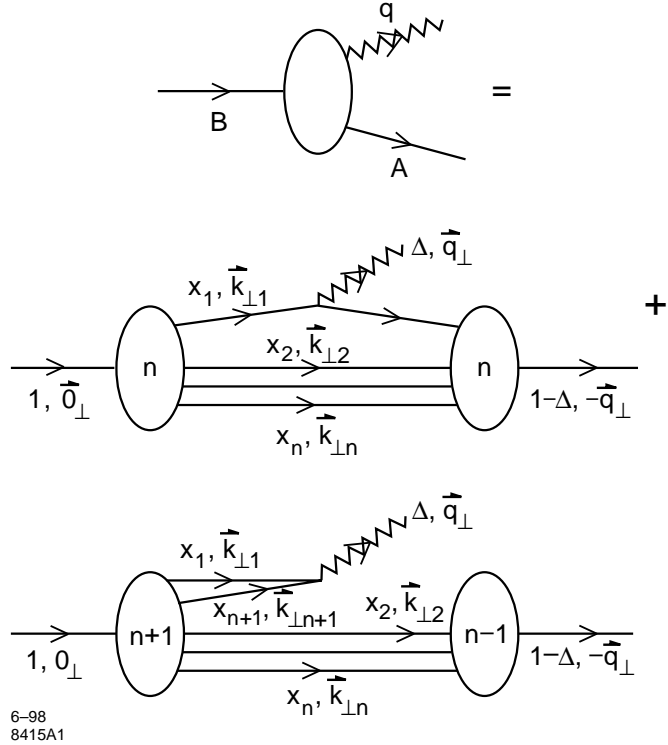}
\end{center}
\caption[*]{Exact representation of electroweak decays and time-like form
factors in the
light-cone Fock representation.
}
\label{fig1}
\end{figure}

Exclusive semi-leptonic $B$-decay amplitudes, such as
$B\rightarrow A \ell \bar{\nu}$ can be evaluated exactly in the light-cone
formalism.\cite{Brodsky:1998hn}
These timelike decay matrix elements require the
computation of
the diagonal matrix element $n \rightarrow n$ where parton number is conserved,
and the off-diagonal $n+1\rightarrow n-1$ convolution where the current
operator
annihilates a $q{\bar{q'}}$ pair in the initial $B$
wavefunction.  (See Fig. \ref{fig1}.)  This term is a consequence of the fact
that the
time-like decay $q^2 = (p_\ell + p_{\bar{\nu}} )^2 > 0$
requires a positive light-cone momentum fraction
$q^+ > 0$.  Conversely for space-like currents, one can choose $q^+=0$, as in
the Drell-Yan-West representation of the space-like electromagnetic form
factors.\cite{DY,BD,West} However, the off-diagonal convolution can yield
a nonzero
$q^+/q^+$ limiting form as $q^+ \rightarrow 0$.  This extra term appears
specifically
in the case of ``bad" currents such as $J^-$ in which the coupling to $q\bar q$
fluctuations in the light-cone wavefunctions are favored.  In effect, the $q^+
\rightarrow 0$ limit generates
$\delta(x)$ contributions as residues of the $n+1\rightarrow n-1$
contributions.
The necessity for  zero mode $\delta(x)$ terms has been noted by
Chang, Root and Yan,\cite{CRY} and Burkardt.\cite{BUR}

The off-diagonal $n+1 \rightarrow n-1$ contributions provide a new
perspective for the physics of $B$-decays.  A semi-leptonic decay involves
not only matrix
elements where a quark changes flavor, but also a contribution where the
leptonic pair is
created from the annihilation of a $q {\bar{q'}}$ pair within the Fock
states of the initial $B$
wavefunction.  The semi-leptonic decay thus can occur from the annihilation of a
nonvalence quark-antiquark pair in the initial hadron.  This feature will carry
over to exclusive hadronic $B$-decays, such as $B^0 \rightarrow
\pi^-D^+$.  In this case the pion can be produced from the coalescence of a
$d\bar u$ pair emerging from the initial higher particle number Fock
wavefunction of the $B$.  The $D$ meson is then formed from the remaining quarks
after the internal exchange of a $W$ boson.

In principle, a precise evaluation of the hadronic matrix elements needed
for $B$-decays
and other exclusive electroweak decay amplitudes requires knowledge of all of
the light-cone Fock wavefunctions of the initial and final state hadrons.
In the case of model gauge theories such as QCD(1+1) \cite{Horn} or collinear
QCD \cite{AD} in one-space and one-time dimensions, the complete evaluation of
the light-cone wavefunction is possible for each baryon or meson bound-state
using the DLCQ method.\cite{DLCQ,AD} It would be interesting to use such
solutions
as a model for physical $B$-decays.

\section{Exclusive Processes in QCD }

Exclusive and diffractive reactions are highly challenging to analyze in QCD
since they require knowledge of the hadron wavefunctions at the amplitude level.
There has been much progress analyzing
exclusive and diffractive reactions at large momentum transfer from first
principles
in QCD.  Rigorous statements can be made on the basis of asymptotic freedom and
factorization theorems which separate the underlying hard quark and gluon
subprocess amplitude from the nonperturbative physics incorporated into the
process-independent hadron distribution amplitudes
$\phi_H(x_i,Q)$,\cite{LB} the valence light-cone
wavefunctions integrated over $k^2_\perp<Q^2$.

In general, hard exclusive hadronic amplitudes such as quarkonium
decay, heavy hadron decay,  and scattering amplitudes where hadrons
are scattered with large momentum transfer can be factorized at leading
power as a
convolution of distribution amplitudes and hard-scattering quark/gluon matrix
elements\cite{LB}
\begin{eqnarray}
\M_{\rm Hadron} &=& \prod_H \sum_n \int
\prod^{n}_{i=1} d^2k_\perp \prod^{n}_{i=1}dx\, \delta
\left(1-\sum^n_{i=1}x_i\right)\, \delta
\left(\sum^n_{i=1} \vec k_{\perp i}\right) \nonumber \\[2ex]
&& \times \psi^{(\lambda)}_{n/H} (x_i,\vec k_{\perp i},\lambda_i)\,
T_H^{(\Lambda)} \ .
\label{eq:e}
\end{eqnarray}
Here $T_H^{(\Lambda)}$ is the underlying quark-gluon
subprocess scattering amplitude in which the (incident and final) hadrons are
replaced by their respective quarks and gluons with momenta $x_ip^+$, $x_i\vec
p_{\perp}+\vec k_{\perp i}$ and invariant mass above the
separation scale $\M^2_n > \Lambda^2$.  At large $Q^2$ one can integrate
over the
transverse momenta.  The leading power behavior of the hard quark-gluon
scattering
amplitude $T_H(\vec k_{\perp i}=0),$ defined for the case where the quarks are
effectively collinear with their respective parent hadron's momentum,
provides the
basic scaling and helicity features of the hadronic amplitude.  The essential
part of the hadron wavefunction is the hadronic distribution amplitudes,
\cite{LB} defined as the integral over transverse momenta of the valence (lowest
particle number) Fock wavefunction, as defined in Eq. \ref{eq:f1}
where the global cutoff $\Lambda$ is identified with the
resolution $Q$.  The distribution amplitude controls leading-twist exclusive
amplitudes at high momentum transfer, and it can be related to the
gauge-invariant Bethe-Salpeter wavefunction at equal light-cone time
$\tau = x^+$.

The $\log Q$ evolution of the hadron distribution amplitudes
$\phi_H (x_i,Q)$ can be derived from the
perturbatively-computable tail of the valence light-cone wavefunction in the
high transverse momentum regime.  The LC ultraviolet
regulators provide a factorization scheme for elastic and inelastic
scattering, separating the hard dynamical contributions with invariant mass
squared $\M^2 > \Lambda^2_{\rm global}$ from the soft physics with
$\M^2 \le \Lambda^2_{\rm global}$ which is incorporated in the
nonperturbative LC wavefunctions.  The DGLAP evolution of quark and gluon
distributions can also be derived in an analogous way by computing the
variation of
the Fock expansion with respect to $\Lambda^2$.
The renormalization scale ambiguities in hard-scattering
amplitudes via commensurate scale
relations\cite{Brodsky:1995eh,Brodsky:1996tb,Brodsky:1999gm} which connect the
couplings entering exclusive amplitudes to the
$\alpha_V$ coupling which controls the QCD heavy quark
potential.\cite{Brodsky:1998dh}

The features of exclusive processes to leading power in the transferred
momenta are well known:

(1) The leading power fall-off is given by dimensional counting rules for
the hard-scattering amplitude: $T_H \sim 1/Q^{n-1}$, where $n$ is the total
number
of fields
(quarks, leptons, or gauge fields) participating in the hard
scattering.\cite{BF,Matveev:1973ra} Thus the reaction is dominated by
subprocesses
and Fock states involving the minimum number of interacting fields.  The
hadronic
amplitude follows this fall-off modulo logarithmic corrections from the
running of
the QCD coupling, and the evolution of the hadron distribution amplitudes.
In some
cases, such as large angle $p p \to p p $ scattering, pinch contributions from
multiple hard-scattering processes must also be
included.\cite{Landshoff:1974ew}
The general success of dimensional counting rules implies that the
effective coupling
$\alpha_V(Q^*)$ controlling the gluon exchange propagators in
$T_H$ are frozen in the infrared, \ie, have an infrared fixed point, since the
effective momentum transfers $Q^*$ exchanged by the gluons are often a
small fraction
of the overall momentum transfer.\cite{Brodsky:1998dh} The pinch contributions
are then suppressed by a factor decreasing faster than a fixed power.\cite{BF}

(2) The leading power dependence is given by hard-scattering amplitudes $T_H$
which conserve quark helicity.\cite{Brodsky:1981kj,Chernyak:1999cj} Since the
convolution of $T_H$ with the light-cone wavefunctions projects out states with
$L_z=0$, the leading hadron amplitudes conserve hadron helicity; \ie, the
sum of
initial and final hadron helicities are conserved.  Hadron helicity conservation
thus follows from the underlying chiral structure of QCD.

(3) Since the convolution of the hard scattering amplitude $T_H$ with the
light-cone
wavefunctions projects out the valence states with small impact parameter,
the essential part of the hadron wavefunction entering a hard exclusive
amplitude has
a small color dipole moment.  This leads to the absence of initial or final
state
interactions among the scattering hadrons as well as the color transparency
of quasi-elastic interactions in a nuclear target.\cite{BM,Frankfurt:1992dx}
Color transparency reflects the underlying gauge theoretic basis
of the strong interactions.
For example, the amplitude for diffractive vector meson photoproduction
$\gamma^*(Q^2) p \to \rho p$, can be written as convolution of the virtual
photon and
the vector meson Fock state light-cone wavefunctions the $g p \to g p$
near-forward matrix element.\cite{Brodsky:1994kf} One can easily show that only
small transverse size $b_\perp \sim 1/Q$ of the vector meson distribution
amplitude is involved.  The sum over the interactions of the exchanged
gluons tend to
cancel reflecting its small color dipole moment.  Since the hadronic
interactions are
minimal,  the
$\gamma^*(Q^2) N \to
\rho N$ reaction at large $Q^2$ can occur coherently throughout a nuclear
target in
reactions without absorption or final state interactions.  The $\gamma^*A
\to V A$ process thus provides a natural framework for testing QCD color
transparency.  Evidence for color transparency in such reactions has been
found by Fermilab experiment E665.\cite{Adams:1997bh}

(4) The evolution
equations for distribution amplitudes which incorporate the operator product
expansion, renormalization group invariance, and conformal symmetry;
\cite{LB,Brodsky:1980ny,Muller:1994hg,Ball:1998ff,Braun:1999te}

(5) Hidden color
degrees of freedom in nuclear wavefunctions reflects the complex color
structure of hadron and nuclear wavefunctions.\cite{bjl83} The hidden color
increases the normalization of nuclear amplitudes such as the deuteron form
factor at large momentum transfer.

The
field of analyzable exclusive processes has recently been expanded to a new
range of QCD processes, such as the highly virtual diffractive processes
$\gamma^* p \to \rho p$,\cite{Brodsky:1994kf,Collins:1997hv} and semi-exclusive
processes such as
$\gamma^* p \to \pi^+ X$ \cite{acw,Brodsky:1998sr,BB} where the $\pi^+$ is
produced in isolation at large $p_T$.  An important new
application of the perturbative QCD analysis of exclusive processes is the
recent analysis of hard $B$ decays such as $B \to \pi \pi$ by Beneke, {\it et
al.}\cite{Beneke:1999br}

\section{The Transition from Soft to Hard Physics}

The existence of an exact formalism
provides a basis for systematic approximations and a control over neglected
terms.  For example, one can analyze exclusive semi-leptonic
$B$-decays which involve hard internal momentum transfer using a
perturbative QCD formalism\cite{BHS,Beneke:1999br} patterned after the analysis
of form
factors at large momentum transfer.\cite{LB} The hard-scattering analysis
proceeds
by writing each hadronic wavefunction as a sum of soft and hard contributions
\begin{equation}
\psi_n = \psi^{{\rm soft}}_n (\M^2_n < \Lambda^2) + \psi^{{\rm hard}}_n
(\M^2_n >\Lambda^2) ,
\end{equation}
where $\M^2_n $ is the invariant mass of the partons in the $n$-particle
Fock state and
$\Lambda$ is the separation scale.
The high internal momentum contributions to the wavefunction $\psi^{{\rm
hard}}_n $ can be calculated systematically from QCD perturbation theory
by iterating the gluon exchange kernel.  The contributions from high
momentum transfer exchange to the
$B$-decay amplitude can then be written as a convolution of a hard-scattering
quark-gluon scattering amplitude $T_H$ with the distribution
amplitudes $\phi(x_i,\Lambda)$, the valence wavefunctions obtained by
integrating the
constituent momenta up to the separation scale
${\cal M}_n < \Lambda < Q$.  This is the basis for the
perturbative hard-scattering analyses.\cite{BHS,Sz,BABR,Beneke:1999br}
In the exact analysis, one can
identify the hard PQCD contribution as well as the soft contribution from
the convolution of the light-cone wavefunctions.
Furthermore, the hard-scattering contribution can be systematically improved.

\section{Measurement of Light-cone Wavefunctions and Tests of Color
Transparency via Diffractive Dissociation.}

Diffractive multi-jet production in heavy
nuclei provides a novel way to measure the shape of the LC Fock
state wavefunctions and test color transparency.  For example, consider the
reaction
\cite{Bertsch,MillerFrankfurtStrikman,Frankfurt:1999tq}
$\pi A \rightarrow {\rm Jet}_1 + {\rm Jet}_2 + A^\prime$
at high energy where the nucleus $A^\prime$ is left intact in its ground
state.  The transverse momenta of the jets have to balance so that
$
\vec k_{\perp i} + \vec k_{\perp 2} = \vec q_\perp < {R^{-1}}_A \ ,
$
and the light-cone longitudinal momentum fractions have to add to
$x_1+x_2 \sim 1$ so that $\Delta p_L < R^{-1}_A$.  The process can
then occur coherently in the nucleus.  Because of color transparency,  \ie,
the cancelation of color interactions in a small-size color-singlet
hadron,  the valence wavefunction of the pion with small impact
separation will penetrate the nucleus with minimal interactions,
diffracting into jet pairs.\cite{Bertsch} The two-gluon exchange process in
effect differentiates the transverse momentum dependence of the hadron's
wavefunction twice.  Thus the $x_1=x$,
$x_2=1-x$ dependence of the di-jet distributions will reflect the shape of
the pion distribution amplitude; the $\vec k_{\perp 1}- \vec k_{\perp 2}$
relative transverse momenta of the jets also gives key information on the
underlying shape of the valence pion
wavefunction.\cite{MillerFrankfurtStrikman,Frankfurt:1999tq} The QCD
analysis can
be confirmed by the observation that the diffractive nuclear amplitude
extrapolated to
$t = 0$ is linear in nuclear number $A$, as predicted by QCD color
transparency.  The integrated diffractive rate should scale as $A^2/R^2_A \sim
A^{4/3}$.  A diffractive dissociation experiment of this type, E791,  is now in
progress at Fermilab using 500 GeV incident pions on nuclear
targets.\cite{E791} The preliminary results from E791 appear to be consistent
with color transparency.  The momentum fraction distribution of the jets is
consistent with a valence light-cone wavefunction of the pion consistent with
the shape of the asymptotic distribution amplitude, $\phi^{\rm asympt}_\pi (x) =
\sqrt 3 f_\pi x(1-x)$.  Data from
CLEO\cite{Gronberg:1998fj} for the
$\gamma
\gamma^* \rightarrow \pi^0$ transition form factor also favor a form for
the pion distribution amplitude close to the asymptotic solution\cite{LB}
to the perturbative QCD evolution
equation.\cite{Kroll,Rad,Brodsky:1998dh,Feldmann:1999wr,Schmedding:1999ap}
It is also possible that the distribution amplitude of the
$\Delta(1232)$ for $J_z = 1/2, 3/2$ is close to the asymptotic form $x_1
x_2 x_3$,  but that the proton distribution amplitude is more complex.
This would explain why the $p \to\Delta$ transition form factor appears to
fall faster at large $Q^2$ than the elastic $p \to p$ and the other $p \to
N^*$ transition form factors.\cite{Stoler:1999nj}
It will thus be very interesting to study diffractive tri-jet production using
proton beams dissociating into three jets on a nuclear target.
$ p A \rightarrow {\rm Jet}_1 + {\rm Jet}_2 + {\rm Jet}_3 + A^\prime $ to
determine the fundamental shape of the 3-quark structure of the valence
light-cone wavefunction of the nucleon at small transverse
separation.\cite{MillerFrankfurtStrikman}

It is also interesting to consider the Coulomb dissociation of hadrons as a
means
to resolve their light-cone wavefunctions.\cite{BHP} In the case of photon
exchange, the transverse momentum dependence of the light-cone wavefunction is
differentiated only once.  For example, consider the process $e p \to
e^\prime {\rm
Jet}_1 + {\rm Jet}_2 + {\rm Jet}_3$ in which the proton dissociates into
three distinct jets at large transverse momentum by scattering on an
electron.  In
the case
of an $e p$ collider such as HERA, one can require all of the hadrons to be
produced outside a forward annular exclusion zone,
$\theta_H >
\theta_{\rm min}$, thus ensuring a minimum transverse momentum of each produced
final
state particle.  The
distribution of hadron longitudinal momentum in each azimuthal sector can
be used
to determine the underlying
$x_1, x_2, x_3$ dependence of the proton's valence three-quark wavefunction.
Such a procedure will allow the proton to self-resolve its fundamental
structure.

One can use incident real and virtual photons:
$ \gamma^* A \rightarrow {\rm Jet}_1 + {\rm Jet}_2 + A^\prime $ to
confirm the shape of the calculable light-cone wavefunction for
transversely-polarized and longitudinally-polarized virtual photons.  At low
transverse momentum, one expects interesting nonperturbative modifications.
Such
experiments will open up a direct window on the amplitude structure of
hadrons at
short distances.

\section {Leading Power Dominance in Exclusive QCD Processes}

As a rule,  exclusive reactions at large momentum transfer
appear to approach the empirical power
law fall-off predicted by dimensional counting.  The PQCD predictions
appear to be
accurate over a large range of momentum transfer, consistent with the small mass
scale of QCD.  These include processes such as the proton form factor, time-like
meson pair production in $e^+ e^-$ and $\gamma
\gamma$ annihilation, large-angle scattering processes such as pion
photoproduction
$\gamma p \to \pi^+ p$, and nuclear processes such as the deuteron form
factor at
large momentum transfer and deuteron photodisintegration.\cite{Brodsky:1976rz} A
spectacular example is the recent measurements at CESR of the photon to pion
transition form factor in the reaction $e \gamma \to e
\pi^0$.\cite{Gronberg:1998fj}
As predicted by leading twist QCD\cite{LB} $Q^2 F_{\gamma
\pi^0}(Q^2)$ is essentially constant for 1 GeV$^2 < Q^2 < 10$ GeV$^2.$
Furthermore, the
normalization is consistent with QCD at NLO if one assumes that the pion
distribution
amplitude takes on the form $\phi^{\rm asympt}_\pi (x) =
\sqrt 3 f_\pi x(1-x)$ which is the asymptotic solution\cite{LB} to the
evolution equation for the pion
distribution amplitude.\cite{Kroll,Rad,Brodsky:1998dh,Schmedding:1999ap}

The measured deuteron form factor and the deuteron photodisintegration
cross section
appear to follow the leading-twist QCD predictions at large momentum
transfers in the
few GeV region.\cite{Holt:1990ze,Bochna:1998ca} The normalization of the
measured deuteron form factor is large compared to model calculations
\cite{Farrar:1991qi} assuming that the deuteron's six-quark wavefunction can be
represented at short distances with the color structure of two color
singlet baryons.
This provides indirect evidence for the presence of hidden color components as
required by PQCD.\cite{bjl83}

There are, however, experimental exceptions to the general success of the
leading twist PQCD approach, such as (a) the dominance of the $J/\psi \to
\rho \pi$
decay which is forbidden by hadron helicity conservation and (b) the strong
normal-normal spin asymmetry $A_{NN}$ observed in polarized elastic $p p
\to p p$
scattering and an apparent breakdown of color transparency at large CM
angles and
$E_{CM} \sim 5$ GeV.  These conflicts with leading-twist PQCD predictions can be
used to identify the presence of new physical effects.  For example,
It is usually assumed that a heavy quarkonium state such as the
$J/\psi$ always decays to light hadrons via the annihilation of its heavy quark
constituents to gluons.  However, the transition $J/\psi \to \rho
\pi$ can also occur by the rearrangement of the $c \bar c$ from the $J/\psi$
into the $\ket{ q \bar q c \bar c}$ intrinsic charm Fock state of the $\rho$ or
$\pi$.\cite{Brodsky:1997fj} On the other hand, the overlap rearrangement
integral in the decay $\psi^\prime \to \rho \pi$ will be suppressed since the
intrinsic charm Fock state radial wavefunction of the light hadrons will
evidently
not have nodes in its radial wavefunction.  This observation provides a natural
explanation of the long-standing puzzle why the $J/\psi$ decays prominently to
two-body pseudoscalar-vector final states, whereas the $\psi^\prime$
does not.  The unusual effects seen in elastic proton-proton scattering at
$E_{CM}
\sim 5$ GeV and large angles could be related to the charm threshold
and the effect of a $\ket{ uud uud c \bar c }$ resonance which would appear
as in
the $J=L=S=1$ $p p $ partial wave.\cite{Brodsky:1988xw}

If the pion distribution amplitude is close to its asymptotic form, then one can
predict the normalization of exclusive amplitudes such as the spacelike
pion form factor $Q^2 F_\pi(Q^2)$.  Next-to-leading order
predictions are available which incorporate higher order
corrections
to the pion distribution amplitude as well as the hard scattering
amplitude.\cite{Muller:1994hg,Melic:1999hg,Szczepaniak:1998sa} However, the
normalization of the PQCD prediction for the pion form factor depends
directly on the
value of the effective coupling
$\alpha_V(Q^*)$ at momenta $Q^{*2} \simeq Q^2/20$.  Assuming
$\alpha_V(Q^*) \simeq 0.4$, the QCD LO prediction appears to be
smaller by approximately a factor of 2 compared to the presently available data
extracted
from the original pion electroproduction experiments from
CEA.\cite{Bebek:1976ww} A
definitive comparison will require a careful extrapolation to the pion pole and
extraction of the longitudinally polarized photon contribution of the $e p
\to \pi^+ n$ data.

Recent experiments at Jefferson laboratory utilizing a new polarization transfer
technique indicate that $G_E(Q^2)/G_M(Q^2)$ falls with increasing momentum
transfer
$-t = Q^2$ in the measured domain $1 < Q^2 < 3 $ GeV$^2$.\cite{Jones:1999rz}
This observation
implies that the helicity-changing Pauli form factor $F_2(Q^2)$ is comparable to
the helicity conserving form factor $F_2(Q^2)$ in this domain.  If such a trend
continues to larger $Q^2$ it would be in severe conflict with the
hadron-helicity conserving principle of perturbative QCD.  If $F_2$ were
comparable to $F_1$ at large $Q^2$ in the case of timelike
processes, such as $p \bar p \to e^+ e^-$, where
$G_E= F_1 + {Q^2\over 4 M_N^2} F_2,$ one would see strong deviations from
the usual $1 + \cos^2{\theta}$ dependence of the differential cross section as
well as PQCD scaling. \cite{Paul}  This seems to be in conflict with the available
data from the $E835$
$\bar p p \to e^+ e^-$ experiment at Fermilab.\cite{Ambrogiani:1999bh}

A debate has continued on whether processes such as the pion
and proton
form factors and elastic Compton scattering $\gamma p \to \gamma p$ might be
dominated by higher twist mechanisms until very large momentum
transfers.\cite{Isgur:1989iw,Radyushkin:1998rt,Bolz:1996sw} For example, if one
assumes that the light-cone wavefunction of the pion has the form
$\psi_{\rm soft}(x,k_\perp) = A \exp (-b {k_\perp^2\over x(1-x)})$, then the
Feynman endpoint contribution to the overlap integral at small $k_\perp$ and
$x \simeq 1$ will dominate the form factor compared to the hard-scattering
contribution until
very large $Q^2$.  However, the above form of $\psi_{\rm soft}(x,k_\perp)$
has no
suppression at $k_\perp =0$ for any $x$; \ie, the
wavefunction in the hadron rest frame does not fall-off at all for $k_\perp
= 0$ and
$k_z \to - \infty$.  Thus such wavefunctions do not represent
soft QCD contributions.  Furthermore, such endpoint contributions will be
suppressed
by the QCD Sudakov form factor, reflecting the fact that a near-on-shell
quark must
radiate
if it absorbs large momentum.  If the endpoint contribution dominates
proton Compton
scattering, then both photons will interact on the same
quark line in a local fashion, and the
amplitude is predicted to be real, in strong contrast to the complex
phase structure of the PQCD predictions.  It should be noted that there is no
apparent endpoint contribution
which could explain the success of dimensional counting ($s^{-7}$ scaling
at fixed
$\theta_{cm}$) in large-angle pion photoproduction.

The perturbative QCD predictions\cite{Kronfeld:1991kp} for the
Compton amplitude phase can be tested in
virtual Compton scattering by interference with Bethe-Heitler
processes.\cite{Brodsky:1972vv} One can also measure the interference of deeply
virtual Compton amplitudes with the timelike form factors by studying reactions
in $e^+ e^-$ colliders such as $e^+ e^- \to \pi^+ \pi^- \gamma$.  The asymmetry
with respect to the electron or positron beam measures the interference of the
Compton diagrams with the amplitude in which the photon is emitted from the
lepton line.

It is interesting to compare the corresponding calculations of form
factors of bound states in QED.  The soft wavefunction
is the Schr\"odinger-Coulomb solution $\psi_{1s}(\vec k) \propto (1 + {{\vec
p}^2/(\alpha m_{\rm red})^2})^{-2}$, and the full wavefunction,  which
incorporates transversely polarized photon exchange, only differs by
a factor $(1 + {\vec p}^2/m^2_{\rm red})$.  Thus the leading twist
dominance of form
factors in QED occurs at relativistic scales $Q^2 > {m^2_{\rm
red}} $.\cite{Brodsky:1989pv}
Furthermore, there are no extra relative factors of $\alpha$ in the
hard-scattering contribution.  If the QCD coupling $\alpha_V$ has
an infrared fixed-point, then the fall-off of the valence wavefunctions of
hadrons will have analogous power-law
forms, consistent with the Abelian correspondence
principle.\cite{Brodsky:1997jk} If such power-law wavefunctions are
indeed applicable to the soft domain of QCD then, the transition to
leading-twist
power law behavior will occur in the nominal hard perturbative QCD domain where
$Q^2 \gg \VEV{k^2_\perp}, m_q^2$.

\section*{Outlook}

It many ways the study of quantum chromodynamics is just beginning.  The
most important features of the theory remain to be solved, such as the
problem of
confinement in QCD, the behavior of the QCD coupling in the infrared, the phase
and vacuum structure/zero mode structure of QCD, the fundamental
understanding of hadronization and parton coalescence at the amplitude
level, and
the nonperturbative structure of hadron wavefunctions.
There are also
still many outstanding phenomenological puzzles in QCD. The precise
interpretation of $CP$ violation and the weak interaction parameters in
exclusive
$B$ decays will require a full
understanding of the QCD physics of hadrons.

Light-cone quantization methods
appear to be especially well suited for progress in understanding the
relevant nonperturbative structure of the theory. Since the Hamiltonian approach
is formulated in Minkowski space, predictions for the hadronic phases needed
for CP violation studies can be obtained. Commensurate scale relations promise a
new level of precision in perturbative QCD predictions which are devoid of
renormalization scale and renormalon ambiguities.
However, progress in QCD is driven by
experiment, and we are fortunate that there are
new experimental facilities such as Jefferson laboratory, the upcoming QCD
studies of exclusive processes $e^+ e^-$ and $\gamma \gamma$ processes at
the high
luminosity
$B$ factories, as well as the new accelerators and colliders now being
planned to
further advance the study of QCD phenomena.

\section*{Acknowledgments}

I have been very fortunate to have had the opportunity
to work in the field of QCD, a theory which is so rich in
opportunities and challenges.  It has been gratifying to be able to work with
outstanding collaborators, and to have guidance by mentors such as Donald
Yennie, T. D. Lee, Sidney Drell, James Bjorken, and Benson Chertok.  I am
grateful to George Strobel for organizing this workshop in my honor and to
all of
my colleagues who participated at the meeting.

\end{document}